# Fermion–Fermion Scattering in the Gross–Neveu Model — a Status Report*†

Meinulf Göckeler[a,b], Hans A. Kastrup[a], Jörg Viola[a,b] and Jörg Westphalen[a,b]

[a]Institut für Theoretische Physik E, RWTH Aachen, D–52056 Aachen, Germany

[b]HLRZ c/o KFA Jülich, P.O.Box 1913, D–52425 Jülich, Germany

Encouraged by the successful applications of Lüscher's method to boson–boson scattering we discuss the possibility of extracting scattering phase shifts from finite–volume energies for fermion–fermion scattering in the Gross–Neveu model.

## 1. INTRODUCTION

Lüscher's method to calculate scattering phase shifts from the two–particle energy spectrum in finite volume has been applied successfully to boson–boson scattering in several models (see e.g. [1] and references therein).

The case of fermion–fermion scattering in the Gross–Neveu model [2] needs two additional considerations. First, we have to extend Lüscher's proofs [3] to the fermion–fermion case. Secondly, one cannot preserve the full flavor symmetry of the continuum Gross–Neveu model on the lattice. Therefore we have to analyze carefully the relation of the lattice energy spectrum to the continuum energy levels.

## 2. LÜSCHER'S RELATION IN d=2

A quantum field theory in finite spatial volume with periodic boundary conditions and extension $L$ has a discrete two–particle energy spectrum $\{W_\nu\}$ or momentum spectrum $\{k_\nu | k_\nu = \sqrt{\frac{1}{4}W_\nu^2 - m^2}\}$. Lüscher's relation [4] in d=1+1 now states that the following equation holds for the spectrum in finite volume and the scattering phase shifts $\delta(k)$ in the continuum

$$\delta(k_\nu) = -\frac{1}{2}k_\nu L \qquad (\mathrm{mod}\,\pi). \tag{1}$$

If the function $\delta(k)$ is known one can solve (1) for $k_\nu$ and calculate the energy spectrum in finite spatial volume $L$.

On the other hand, suppose one has calculated, for example by Monte–Carlo methods, some of the lower energy levels in finite volume. Then one can use (1) to calculate the scattering phase shifts at the corresponding momenta *directly*, that means without an extrapolation to $L \to \infty$.[1]

Note that Lüscher considered (1) in the framework of boson–boson scattering. Nevertheless (1) can also be applied to fermion–fermion scattering, as one can generalize Lüscher's proofs in [3] to the fermionic case.

## 3. THE GROSS–NEVEU MODEL

The Gross–Neveu model [2] with $n_f$ flavors has the following euclidean action:

$$S_c = \int d^2x \left\{ \overline{\Psi}^a \gamma_\mu \partial_\mu \Psi^a - \frac{g^2}{2}\left(\overline{\Psi}^a \Psi^a\right)^2 \right\} , \tag{2}$$

where $a = 1, \ldots, n_f$. As is well known already from [2], $S_c$ has not only the obvious $U(n_f)$ flavor symmetry but an $O(2n_f)$ invariance, which becomes apparent by introducing the fields

$$\Psi^{a,1}(x) = \tfrac{1}{2}\left(\Psi^a(x) + {}^c\Psi^a(x)\right) ,$$
$$\Psi^{a,2}(x) = \tfrac{1}{2i}\left(\Psi^a(x) - {}^c\Psi^a(x)\right) ,$$

[1]This is only partially true, since (1) is only valid provided polarization effects, which are of order $e^{-mL}$, are negligible [3,1].

*supported by the *Deutsche Forschungsgemeinschaft*
†talk given by Jörg Westphalen at Lattice 94

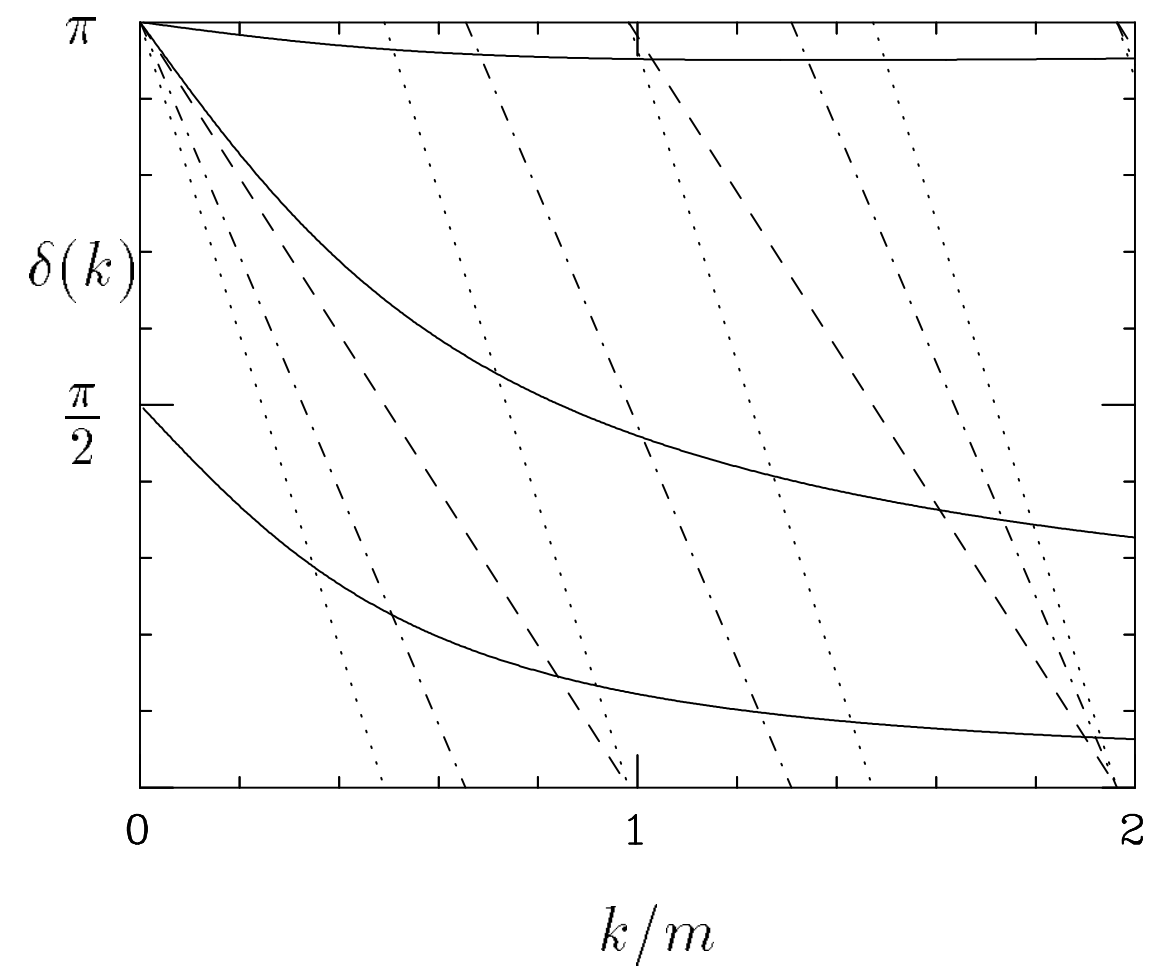

Figure 1. Scattering phase shifts in the Gross–Neveu model for $n_f = 4$ and $m = 0.2$. The upper, middle and lower solid curve show the scattering phase shifts $\delta_{sym}$, $\delta_{inv}$ and $\delta_{as}$, respectively. The dashed, dot dashed and dotted lines show the r.h.s. of (1) for $L = 32$, 48 and 64.

where ${}^c\Psi^a = C\overline{\Psi}^{aT}$ is the charge–conjugate field. The action now reads:

$$S_c = -\int d^2x \left\{ \Psi^{a,iT} C^{-1} \gamma_\mu \partial_\mu \Psi^{a,i} + \frac{g^2}{2} \left( \Psi^{a,iT} C^{-1} \Psi^{a,i} \right)^2 \right\}, \quad (3)$$

where $a = 1, \ldots, n_f$, $i = 1, 2$. Besides the usual space–time symmetries, $S_c$ is invariant under charge conjugation, which is now part of the $O(2n_f)$, and under a discrete chiral transformation called $\gamma_5$ [2].

Since the fermions transform according to the vector representation of $O(2n_f)$, the S–matrix as well as the two–particle energy spectrum decomposes into three sectors (corresponding to invariant, traceless symmetric and antisymmetric tensors of rank two) and Lüscher's relation holds for each sector separately [4]. Due to conformal invariance the three phase shifts $\delta_{inv}(k)$, $\delta_{sym}(k)$ and $\delta_{as}(k)$ can be calculated analytically [5,6]. Thus in a first step we can use (1) to estimate the finite–volume energies for some typical values of $L$. From fig. 1 we see that there should be a sufficient number of energy levels in the low–energy regime for numerically realizable lattices (i.e. for $L \leq 64$).

## 4. THE GROSS–NEVEU MODEL ON THE LATTICE

As a lattice version of the two–dimensional Gross–Neveu model we use $N$ staggered fermions with "hypercubic" interaction:

$$S = \sum_{x,\mu} \frac{1}{2} \eta_\mu(x) \left( \overline{\chi}^f_x \chi^f_{x+\mu} - \overline{\chi}^f_{x+\mu} \chi^f_x \right) - \frac{g^2}{2} \sum_x \left( \sum_\rho \overline{\chi}^f_{x+\rho} \chi^f_{x+\rho} \right)^2 , \quad (4)$$

where $f = 1, \ldots, N$. This model describes $n_f = 2N$ massive fermions. Again the flavor symmetry of the action is larger than the obvious $U(N)$. By means of the new fields

$$\chi^{f,1}(x) = \tfrac{1}{2} \left( \chi^f(x) + {}^c\chi^f(x) \right) ,$$
$$\chi^{f,2}(x) = \tfrac{1}{2i} \left( \chi^f(x) - {}^c\chi^f(x) \right) ,$$

where ${}^c\chi^f(x) = \epsilon(x)\overline{\chi}^f(x)$ is the lattice charge conjugate field, the action can be written as:

$$S = \sum_{x,\mu} \frac{i}{2} \eta_\mu(x) \epsilon(x) \left( \chi^{f,i}_{x+\mu} \begin{pmatrix} 0 & 1 \\ -1 & 0 \end{pmatrix}_{ij} \chi^{f,j}_x \right) - \frac{g^2}{2} \sum_x \left( \sum_\rho \epsilon(x+\rho) \chi^{f,i}_{x+\rho} \begin{pmatrix} 0 & 1 \\ -1 & 0 \end{pmatrix}_{ij} \chi^{f,j}_{x+\rho} \right)^2 .$$

$S$ is invariant under the group *USp(N)* of unitary symplectic transformations in $2N$ dimensions.

Hence the lattice flavor symmetries form only a subgroup of the continuum flavor group. The relation of the two symmetry groups is determined by the (naive) continuum limit.

## 5. CALCULATION OF THE ENERGY SPECTRUM

We shall calculate the finite–volume energies on the lattice from connected correlation functions of operators $\mathcal{O}_i(t)$ localized in time:

$$\mathcal{C}_{ij}(\tau) = \langle \mathcal{O}_i(\tau)^* \mathcal{O}_j(0) \rangle_c .$$

Since we are interested in the energy spectrum in a specific symmetry sector we choose the $\mathcal{O}_i$ to form a bosonic momentum zero irreducible representation (irrep) of the lattice time–slice group (LTS), which is the subgroup of the symmetry group of $S$ leaving the time slices invariant. The group LTS is generated by the spatial inversion $I$, the lattice version $A$ of $\gamma_5$, the spatial shift $S_1$, and the *USp(N)* transformations $M$.

The bosonic zero momentum irreps of LTS can be found easily (like in [7]) to be

$$\Delta_D^{\sigma_I\sigma_\epsilon\sigma_1}:$$
$$I^\alpha A^\beta S_1^\gamma M \longmapsto \sigma_I^\alpha \sigma_\epsilon^\beta \sigma_1^\gamma\, D(\begin{pmatrix} 0_N & 1_N \\ -1_N & 0_N \end{pmatrix}^\gamma M).$$

Here the $\sigma$'s are $\pm 1$ and $D$ is some irrep of *USp(N)* on tensors of even rank. On the lattice we therefore characterize the symmetry sectors by $D$, $\sigma_I$, $\sigma_\epsilon$ and $\sigma_1$.

By the naive continuum limit the lattice time slice group is embedded into the continuum time slice group (CTS), i.e. LTS is a subgroup of CTS. CTS is generated by the spatial translations, the parity transformation $P$, the $\gamma_5$ transformation, and the $O(4N)$ transformations $\overline{M}$. The bosonic momentum zero irreps of CTS are:

$$\overline{\Delta}_{\overline{D}}^{\sigma_p\sigma_5}:$$
$$P^\alpha \gamma_5^\beta \overline{M} \longmapsto \sigma_p^\alpha \sigma_5^\beta\, \overline{D}(\begin{pmatrix} 0_{2N} & 1_{2N} \\ -1_{2N} & 0_{2N} \end{pmatrix}^{\alpha+\beta} \overline{M}).$$

Again the $\sigma$'s are $\pm 1$ and $\overline{D}$ is some irrep of $O(4N)$ on tensors of even rank.

Since LTS is a subgroup of CTS, $\overline{\Delta}_{\overline{D}}^{\sigma_p\sigma_5}$ induces a representation of LTS, which may be reducible:

$$\overline{\Delta}_{\overline{D}}^{\sigma_p\sigma_5}\downarrow^{LTS} = \oplus \Delta_D^{\sigma_I\sigma_\epsilon\sigma_1}.$$

We calculated this decomposition explicitly for the case we are especially interested in where $\overline{D}$ is an irrep on tensors of rank two. For example:

$$\overline{\Delta}_{sym}^{++}\downarrow^{LTS} = \Delta_{(2,0,\ldots)}^{-+-} \oplus \Delta_{(1,1,0,\ldots)}^{+-+} \oplus \Delta_{(2,0,\ldots)}^{++-} \oplus \Delta_{(2,0,\ldots)}^{--+}.$$

For the definition of the irreps $(\lambda_1, \ldots, \lambda_N)$ of *USp(N)* see [8]. From the full table given on the last page we find that each two–particle operator $\mathcal{O}_i$ belonging to some representation $\Delta_D^{\sigma_I\sigma_\epsilon\sigma_1}$ couples to two continuum symmetry sectors (i.e. two $\overline{\Delta}_{\overline{D}}^{\sigma_p\sigma_5}$) with *different* $\sigma_p$. Since the corresponding energies $E$ and $E'$ may be distinguished by their different contributions $e^{-E\tau}$ and $(-)^\tau e^{-E'\tau}$ to the correlation matrix $\mathcal{C}_{ij}$, we can identify for each energy level the corresponding *continuum* symmetry sector. We then apply Lüscher's relation in each sector separately.

## 6. SUMMARY AND OUTLOOK

Lüscher's method for calculating elastic scattering phase shifts non–perturbatively, which proved successful for boson–boson scattering in several models, has been generalized such that it can be applied to fermion–fermion scattering in the two–dimensional Gross–Neveu model. So we are now in a position to perform numerical simulations, calculate correlation functions and extract the phase shifts, which can be compared with the analytical results. These computations are currently under way.

| $\overline{\Delta}^{\sigma_p\sigma_5}_{\bar{D}}$ | | | $\Delta^{\sigma_1\sigma_I\sigma_\epsilon}_{D}$ | | | | | | | |
|---|---|---|---|---|---|---|---|---|---|---|
| $\bar{D}$ | $\sigma_p$ | $\sigma_5$ | $D$ | $\sigma_1$ | $\sigma_I$ | $\sigma_\epsilon$ | $D$ | $\sigma_1$ | $\sigma_I$ | $\sigma_\epsilon$ |
| **sym.** | + | + | $(2,0,0,\ldots,0)$ | − | − | + | | | | |
| | | | $(1,1,0,\ldots,0)$ | + | + | − | | | | |
| | | | $(2,0,0,\ldots,0)$ | − | + | + | | | | |
| | | | $(2,0,0,\ldots,0)$ | + | − | − | | | | |
| | + | − | $(2,0,0,\ldots,0)$ | − | − | − | | | | |
| | | | $(1,1,0,\ldots,0)$ | + | + | + | | | | |
| | | | $(2,0,0,\ldots,0)$ | − | + | − | | | | |
| | | | $(2,0,0,\ldots,0)$ | + | − | + | | | | |
| | − | + | $(2,0,0,\ldots,0)$ | − | + | + | | | | |
| | | | $(1,1,0,\ldots,0)$ | + | − | − | | | | |
| | | | $(2,0,0,\ldots,0)$ | − | − | + | | | | |
| | | | $(2,0,0,\ldots,0)$ | + | + | − | | | | |
| | − | − | $(2,0,0,\ldots,0)$ | − | + | − | | | | |
| | | | $(1,1,0,\ldots,0)$ | + | − | + | | | | |
| | | | $(2,0,0,\ldots,0)$ | − | − | − | | | | |
| | | | $(2,0,0,\ldots,0)$ | + | + | + | | | | |
| **antisym.** | + | + | $(1,1,0,\ldots,0)$ | − | − | + | $(0,\ldots,0)$ | − | − | + |
| | | | $(2,0,0,\ldots,0)$ | + | + | − | | | | |
| | | | $(1,1,0,\ldots,0)$ | − | + | + | $(0,\ldots,0)$ | − | + | + |
| | | | $(1,1,0,\ldots,0)$ | + | − | − | $(0,\ldots,0)$ | + | − | − |
| | + | − | $(1,1,0,\ldots,0)$ | − | − | − | $(0,\ldots,0)$ | − | − | − |
| | | | $(2,0,0,\ldots,0)$ | + | + | + | | | | |
| | | | $(1,1,0,\ldots,0)$ | − | + | − | $(0,\ldots,0)$ | − | + | − |
| | | | $(1,1,0,\ldots,0)$ | + | − | + | $(0,\ldots,0)$ | + | − | + |
| | − | + | $(1,1,0,\ldots,0)$ | − | + | + | $(0,\ldots,0)$ | − | + | + |
| | | | $(2,0,0,\ldots,0)$ | + | − | − | | | | |
| | | | $(1,1,0,\ldots,0)$ | − | − | + | $(0,\ldots,0)$ | − | − | + |
| | | | $(1,1,0,\ldots,0)$ | + | + | − | $(0,\ldots,0)$ | + | + | − |
| | − | − | $(1,1,0,\ldots,0)$ | − | + | − | $(0,\ldots,0)$ | − | + | − |
| | | | $(2,0,0,\ldots,0)$ | + | − | + | | | | |
| | | | $(1,1,0,\ldots,0)$ | − | − | − | $(0,\ldots,0)$ | − | − | − |
| | | | $(1,1,0,\ldots,0)$ | + | + | + | $(0,\ldots,0)$ | + | + | + |
| **inv.** | + | + | | | | | $(0,\ldots,0)$ | + | + | + |
| | + | − | | | | | $(0,\ldots,0)$ | + | + | − |
| | − | + | | | | | $(0,\ldots,0)$ | + | − | + |
| | − | − | | | | | $(0,\ldots,0)$ | + | − | − |

Table 1
Decomposition of $\overline{\Delta}^{\sigma_p\sigma_5}_{\bar{D}}$ with respect to LTS for $N \leq 2$. For the definition of the irreps $(\lambda_1,\ldots,\lambda_N)$ of *USp(N)* see [8].